# Atomic scale understanding of the role of hydrogen and oxygen segregation in the embrittlement of grain boundaries in a twinning induced plasticity steel

Heena Khanchandani[1], Baptiste Gault[1,2*]

[1]Department of Microstructure Physics and Alloy Design, Max-Planck-Institut für Eisenforschung, Max-Planck-Str. 1, 40237 Düsseldorf, Germany.

[2]Department of Materials, Royal School of Mines, Imperial College, Prince Consort Road, London SW7 2BP, United Kingdom.

[*]Corresponding author: b.gault@mpie.de

**Abstract:** High strength twinning induced plasticity (TWIP) steels have potential for applications in the automotive industry. However, they are prone to hydrogen embrittlement (HE) and galvanic corrosion. Here, we report on the susceptibility towards HE and oxidation of a model Fe 27Mn 0.3C (wt%) TWIP steel by atom probe tomography. We measured the segregation of hydrogen and oxygen at grain boundaries, which appears correlated with manganese depletion. Our study suggests a correlation between HE and oxidation mechanisms in TWIP steels, which can combine to favor the previously reported hydrogen enhanced decohesion of grain boundaries.

High manganese austenitic, i.e. face centered cubic, twinning induced plasticity (TWIP) steels are promising high strength metallic materials for automotive applications [1]. However, they are highly susceptible to hydrogen embrittlement (HE) [2] and corrosion [3]. Intergranular fracture in these materials has been reported upon their exposure to hydrogen and moisture containing environments [2,4]. It has been proposed that intergranular fracture is pronounced at the locations with higher hydrogen contents [5]. Hydrogen diffusion is accelerated at grain boundaries (GBs) [6,7] which are also typically the locations where hydrogen is found segregated, making the material more susceptible to intergranular cracking. It is hence essential to understand the local chemistry at GBs to understand the underlying mechanisms.

Although higher manganese (above 20 wt%) is sometimes added to TWIP steels in order to achieve a stable austenitic structure at room temperature [8], it has been suggested that their high susceptibility to HE and galvanic corrosion is due to their higher manganese content [3,9]. Indeed, a high manganese content increases the solubility and the mobility of hydrogen in high manganese steels [9], making more diffusible hydrogen available [10] which can further facilitate cracking. The corrosion resistance of TWIP steels has been examined and a GB engineering approach [11] has been proposed to improve their resistance to intergranular corrosion [12] and HE [13]. However, limited studies have correlated the mechanisms for HE and corrosion to GB chemistries. Based on a recent preliminary study [14], there are indications that the role of oxygen may have been underestimated.

Here, we studied the segregation of oxygen and deuterium at high-angle GBs by atom probe tomography (APT) in a model TWIP steel to better understand these mechanisms at the atomic scale. Electron backscatter diffraction (EBSD) and transmission Kikuchi diffraction (TKD) were performed to first select the GBs and determine their misorientation according to Brandon's criterion [15]. To facilitate the quantification of hydrogen by APT, we used isotope labelling with deuterium, as discussed in ref. [16–18]. The deuterium gas charging and cryogenic transfer workflows were performed using the infrastructure described in ref. [18,19]. Oxygen and deuterium were found to be segregated at GBs in our study, which was also accompanied by a strong manganese depletion. We also observed that oxygen could not diffuse into the GBs decorated with carbon, which can help to design the steels resistant to oxidation and HE.

We selected a model Fe 27Mn 0.3C (wt%) TWIP steel. The strip cast steel was homogenized at 1150ºC for 2 hours and several hundreds of microns were cut from the surface of the sample to remove the surface oxide regions. Subsequently, it was cold-rolled to achieve a 50% thickness reduction. The material was then subjected to a recrystallization annealing treatment at 800ºC for 20 minutes, followed by water cooling to room temperature. The bulk chemical composition (at%) of the studied TWIP steel is listed in Table 1.

*Table 1. Bulk chemical composition of the studied TWIP steel*

| Elements | Mn | C    | O    | Fe      |
|----------|----|------|------|---------|
| at%      | 27 | 1.27 | 0.08 | Balance |

EBSD-orientation mapping was performed using a Zeiss Sigma 500 SEM equipped with an EDAX/TSL Hikari camera at an accelerating voltage of 15kV, a beam current of 9nA, a scan step size of 0.5µm, a specimen tilt angle of 70º, and a working distance of 14mm [20]. FEI Helios NanoLab 600i dual-beam FIB/SEM instrument equipped with a Hikari camera was used for preparing the APT specimens containing the GB of interest by using the standard site-specific lift-out procedure [21]. TKD was performed on APT specimens with a step size of 20nm. EBSD and TKD analyses were carried out using OIM Data Analysis 7.0.1 (EDAX Inc.) software. APT experiments were performed in LEAP 5000XS and XR instruments (CAMECA Instruments Inc. Madison, WI, USA) in voltage pulsing mode at 15-20% pulse fraction and 200kHz pulse repetition rate. The specimen's base temperature was kept at 70K and the target detection rate was set to 5 ions every 1000 pulses. These analysis conditions have been determined for the reliable analysis of hydrogen in a previous study on the same material [17].

Figure 1 summarizes the APT analysis of a GB from the uncharged, recrystallized TWIP steel, with a misorientation angle of ~50° determined by TKD in Figure 1a. The 3-D elemental map following the APT reconstruction in Figure 1b shows iron, manganese, carbon and oxygen atoms and the 1-D composition profile across the grain boundary (blue arrow) is plotted in Figure 1c. Carbon segregates up to 4.11 ± 0.2 at% accompanied by a slight manganese segregation of 27.6 ± 0.5 at%. The chemical width of the GB was determined by the full width at half maximum of the peak of the profile indicated by the dotted lines in Figure 1c. It is approx. 2 nm for manganese and ~3.2 nm for carbon.

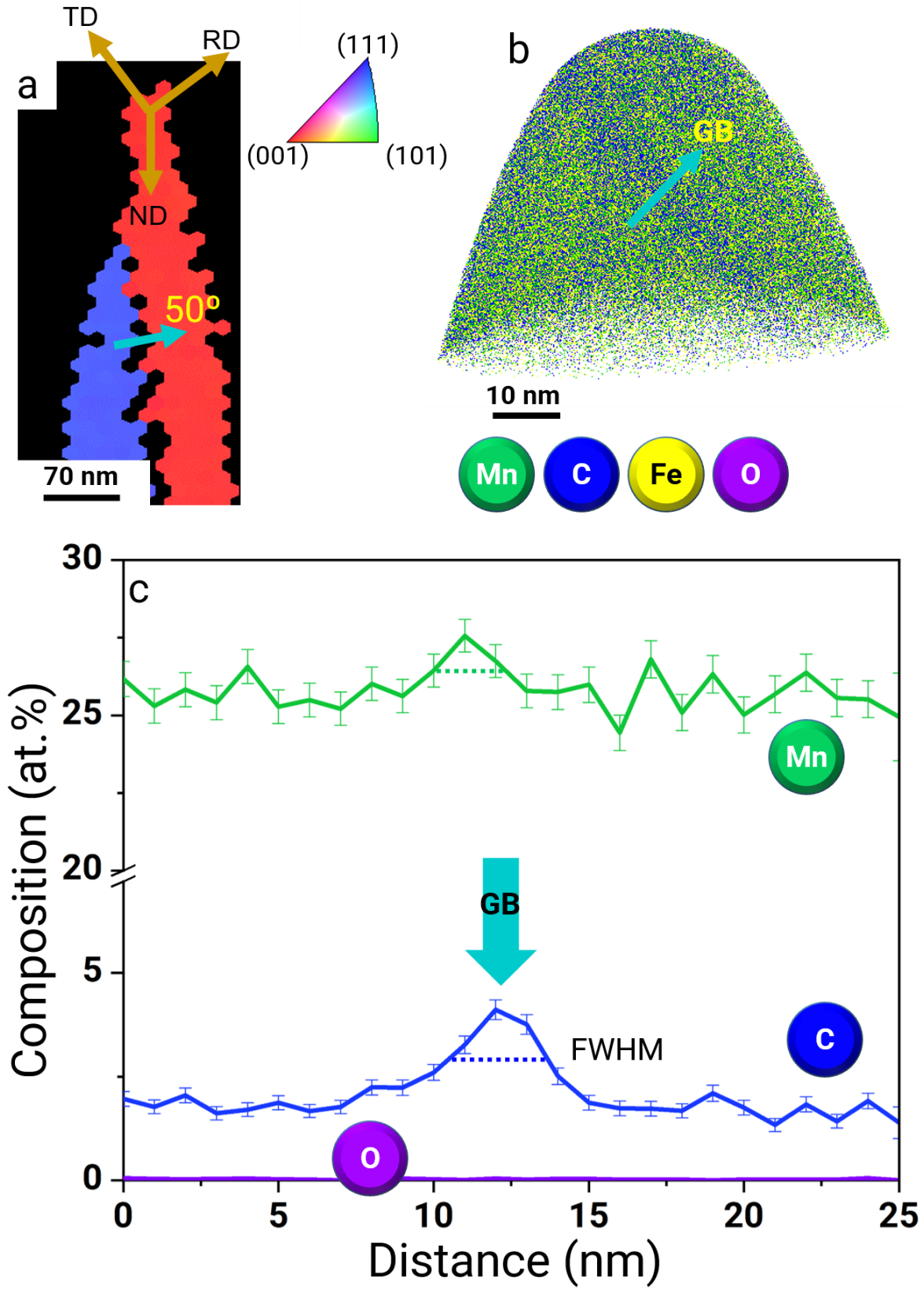

*Figure 1. (a) Transmission Kikuchi Diffraction – inverse pole figure (TKD-IPF) map along out of the plane direction of the studied grain boundary (GB). ND – normal direction in the plane, TD – transverse direction, RD – rolling direction. (b) The 3-D elemental map showing iron, carbon,*

*manganese and oxygen atoms. (c) The 1-D composition profile calculated across the GB with 1 nm bin width. The dotted lines indicate the chemical width of GB determined by full width at half maximum (FWHM) of the profile.*

Another GB from this recrystallized sample was analyzed by APT. The TKD-inverse pole figure (IPF) map of the specimen shown in Figure 2a reveals two GBs with misorientation angles of ~18º and ~46º. The specimen was first analyzed by APT until the first GB appeared in the field-of-view, as shown in supplementary Figure S1a, at which point the measurement was interrupted. The mass spectrum in the supplementary Figure S1b contains only a peak at 1Da corresponding to the background hydrogen signal [17,18], while no peaks at 2 or 3Da were observed.

Subsequently, the specimen was transferred to a gas charging chamber described in Ref. [17,18], in a ultra-high vacuum (UHV) suitcase [19]. The specimen was exposed to 250 mbar of deuterium ($D_2$) gas atmosphere at room temperature for 6 hours. After charging, the specimen was transferred to the atom probe for analysis through the UHV suitcase precooled to liquid nitrogen temperature. The reconstructed elemental map from the analysis of the specimen after charging is displayed in Figure 2b. The mass spectrum shown in supplementary Figure S1c evidences successful deuterium charging of the specimen, by exhibiting the peaks at 2 and 3Da, corresponding to $D^+$ and $HD^+$.

The GBs are highlighted by the segregation of deuterium and oxygen in Figure 2b, and appear curved, with indications of faceting [22–29]. 1-D composition profiles were calculated across different facets of both GBs, and are plotted in Figure 2(c-g). The oxygen and deuterium are segregated to all facets, while manganese is depleted. The point density profiles are plotted in supplementary Figures S2(a-b) for two of the selected composition profiles, along with the Mn:Fe point densities (Figures S2(c-d)), which support the depletion of manganese at the GB, accompanied by the segregation of oxygen and deuterium. The local chemical width of the GB is estimated by the full-width-half-minimum (FWHM) for manganese and the full-width-half-maximum (FWHM) for oxygen and deuterium, which are displayed as dotted lines in Figure 2(c-g). The manganese depletion width is narrower than the oxygen and deuterium segregation widths for most of the profiles. The FWHM of each of these profiles is plotted in supplementary Figure S3.

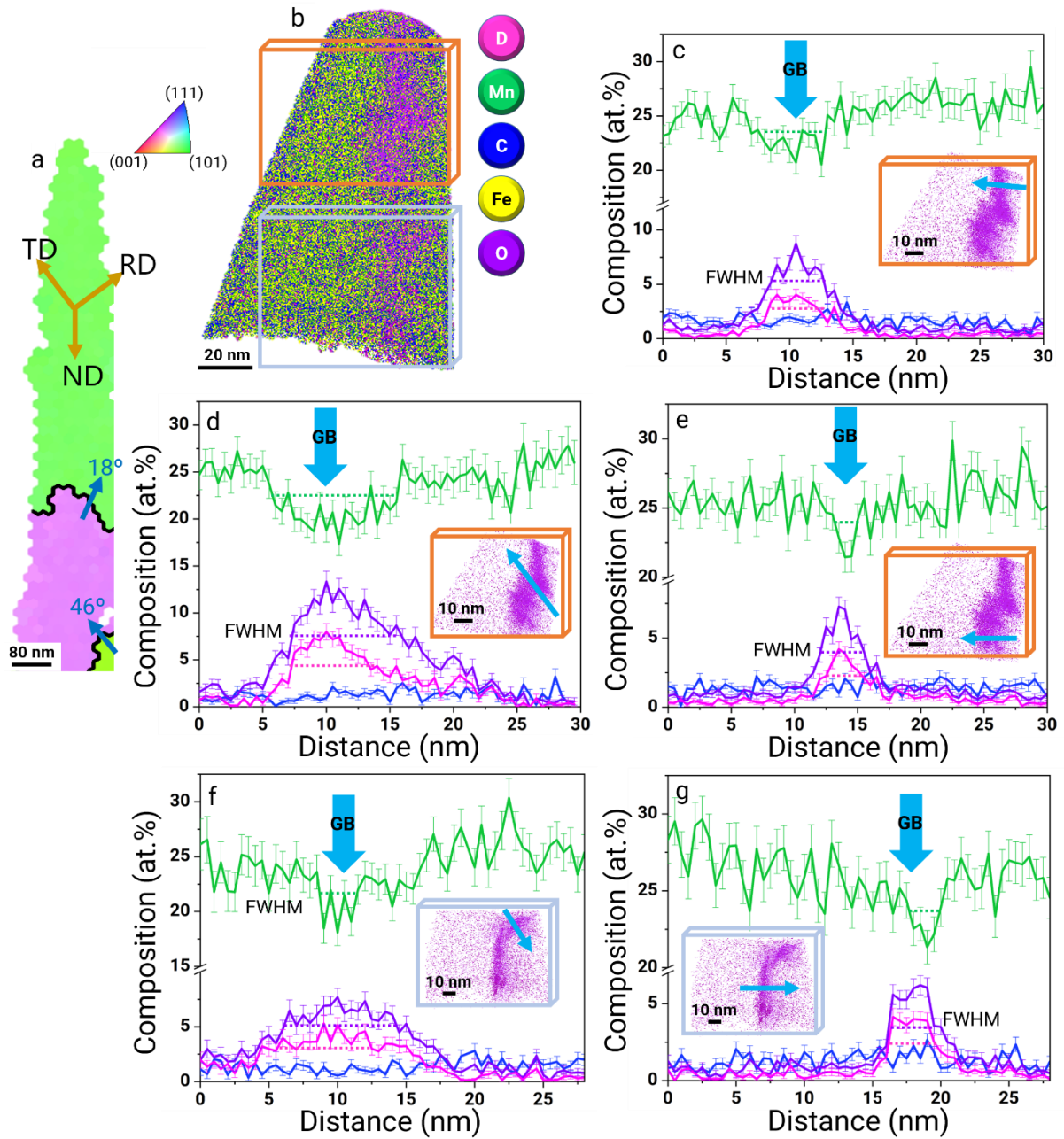

*Figure 2. (a) TKD-IPF map of the APT specimen along out of the plane direction exhibiting two GBs. ND – normal direction in the plane, TD – transverse direction, RD – rolling direction. (b) The 3-D elemental map of the specimen following the deuterium charging, which shows iron, carbon, manganese, oxygen and deuterium atoms. (c-g) The composition profiles calculated with 0.5 nm bin width across different facets of the two GBs.*

An uncharged tensile specimen of the recrystallized TWIP steel sample with gauge dimensions of 4×2×1 mm³ was deformed to fracture. The total elongation to fracture was 75% with the yield strength of 250 MPa and the ultimate tensile strength of ~700 MPa – see ref [30] for further details. A GB with a misorientation angle of ~57° was selected for APT analysis in the deformed sample at a distance of approx. 1mm from the fracture surface. It is highlighted by a blue line in the EBSD-IPF map in Figure 3a. Following deuterium charging, the APT analysis was performed. The reconstructed APT 3-D elemental map is shown in Figure 3b. The local change in point density and pattern in the detector hit map (Figure S4a) also supports the presence of GB. A 1-D composition profile was calculated across the GB in Figure 3c which evidences a carbon segregation of 3.75 ± 0.4 at%, a deuterium segregation of 4.36 ± 0.4 at% and an oxygen segregation of 3.2 ± 0.4 at% at the GB. It also exhibits a manganese depletion by 3 ± 0.9 at%. The large width of GB can be ascribed to the GB migration associated to the deformation [31]. Supplementary Figure S4b plots the corresponding mass spectrum in which the peak at 4 ($D_2^+$) evidences the deuterium ingress into the specimen. The specimen's surface is enriched in oxygen and deuterium, as quantified by a 1-D composition profile calculated in a cylindrical region of interest positioned from the surface through the GB over the length of approx. 25 nm enclosing the GB, and with a diameter of 10 nm in Figure 3d, in the direction indicated by the blue arrow in Figure 3b. Oxygen reaches 16 ± 0.78 at% and deuterium 9.4 ± 0.62 at% at the surface. This surface oxidation illustrates that the surface first corrodes, leading to the ingress of both oxygen and hydrogen into the GB, which can then promote intergranular fracture [30], likely through the decohesion of GBs described by the hydrogen-enhanced decohesion (HEDE) mechanism [32].

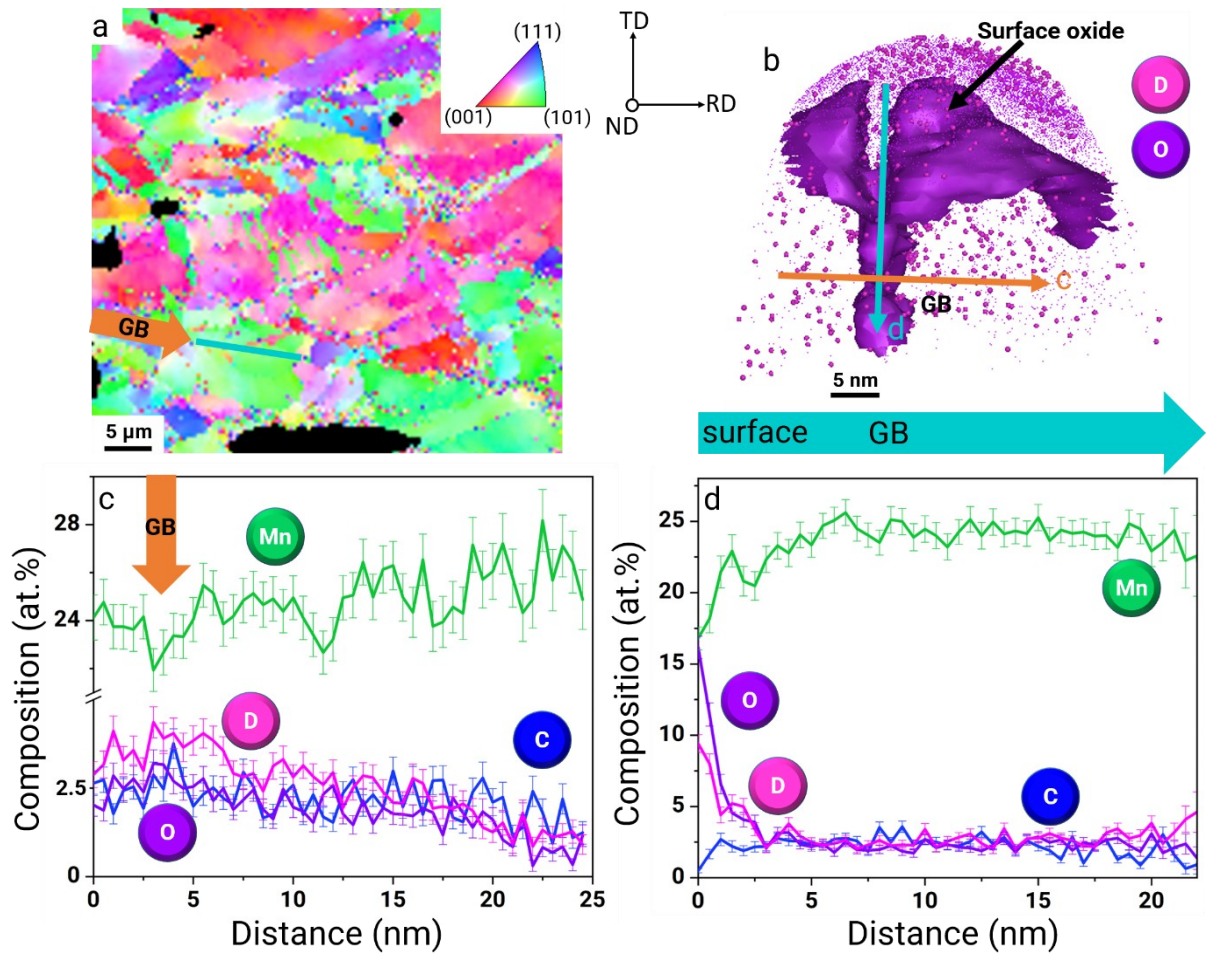

*Figure 3. (a) EBSD-IPF map with respect to the normal direction highlighting the studied GB in blue. ND – normal direction out of the plane, TD – transverse direction, RD – rolling direction. (b) The 3-D elemental map highlighting the GB by violet isosurfaces that encompass the regions containing more than 3.9 at.% of oxygen. (c) The 1-D composition profile calculated across the GB with 0.5 nm bin width. (d) The 1-D composition profile calculated from the surface through the GB with 0.5 nm bin*

*width. The composition profile in (d) is calculated in the direction perpendicular to the profile shown in (c).*

A significant, although unforeseen oxygen segregation was evidenced in the deuterium charged specimens. We already suspected that oxygen can combine with hydrogen to embrittle twin boundaries in TWIP steels [14]. This oxygen could originate from residuals from the fabrication of the alloy. The oxygen content of the alloy was measured in a reduction fusion in a helium atmosphere and revealed that the alloy contains 0.08 at% of oxygen, which could have segregated to GBs during the recrystallization annealing treatment. The oxygen may also have been introduced during charging and transfer of the APT specimen, in particular moisture present along the cryogenic chain [17] could form frost on the specimen's surface and cause corrosion, albeit unlikely at such low temperatures.

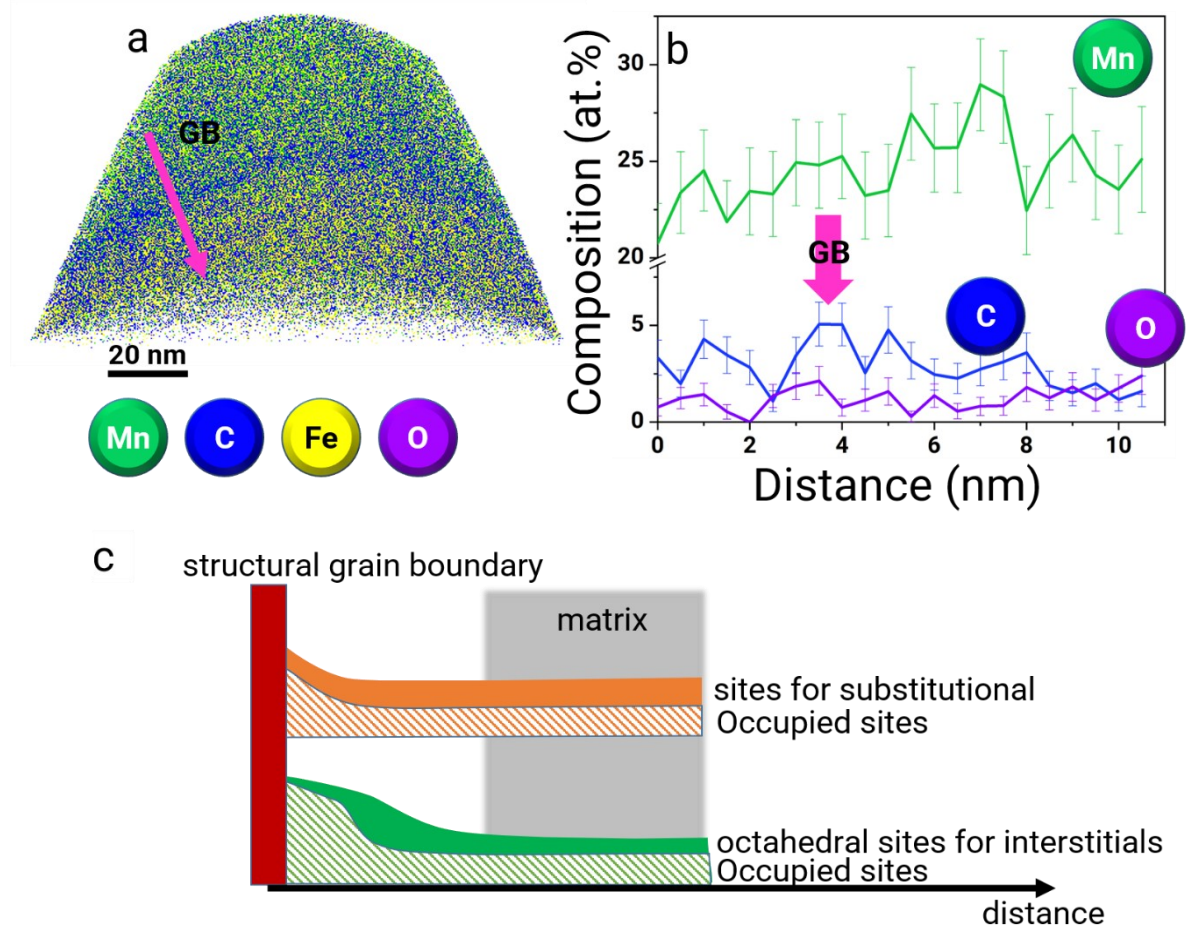

*Figure 4. (a) The 3-D elemental map of the GB highlighted by the carbon segregation. (b) The 1-D composition profile calculated across the GB with 0.5 nm bin width. (c) Schematic illustrating the occupation of substitutional and interstitial sites as a function of distance to the GB.*

Considering that carbon is an interstitial element in austenite and the octahedral interstitial site is energetically favorable for carbon, oxygen and hydrogen in fcc-Fe [32–35], if the octahedral interstitial sites are already occupied by carbon atoms, which improve the GB cohesion [36,37], then it prevents the oxygen and likely hydrogen ingress into the GB and protects it from embrittlement [38]. Figure S5a compiles the GB composition in oxygen and deuterium as a function of carbon's for the GBs analyzed herein. It shows a loose downward trend, suggesting that a stronger carbon segregation hinders the diffusion of oxygen and hydrogen to GBs, which agrees with previous reports [28,36]. The measured compositions of both oxygen and deuterium are plotted as a function of the composition of manganese for each of the profiles reported here, in supplementary Figure S5b, also shows a trend for higher segregation in regions with a relatively lower manganese content.

The formation enthalpy of carbon at the interstitial octahedral site in fcc-Fe is -9.1 eV and that of oxygen occupying the same site is -5.5 eV [33], it is hence favorable for both carbon and oxygen to reside in the octahedral interstitial site. However, the energy required to replace a carbon atom by an oxygen atom is +3.6 eV which is unfavorable, hence the octahedral site occupied by a carbon atom cannot be favorably replaced by an oxygen atom which potentially makes the GBs decorated with carbon more resistant to oxidation. To further support this observation, we exposed a GB decorated with carbon, shown in Figure 4a, to an oxygen ($O_2$) gas atmosphere of 250 mbar for 2 hours at room temperature and analyzed it by APT. The local change in point density and pattern in the detector event map in Figure S6 also supports the presence of GB in the analyzed APT dataset. However, no oxygen enrichment was found at the GB as shown by a 1-D composition profile calculated across the GB in Figure 4b, where 5 ± 1.1 at% of carbon is segregated.

GBs showing no carbon segregation along with manganese depletion are likely to be enriched in iron, making these GBs more susceptible to corrosion [39]. It has been reported that the oxidation of iron can take place even at room temperature [40]. Supplementary Figure S7a plots the Fe:Mn composition ratio as a function of the sum of oxygen and deuterium compositions for each of the analyzed GBs and it also shows a loose trend for higher oxygen and deuterium contents in the regions relatively enriched in iron. Analyses in Figures S5b and S7b suggest that GBs depleted in manganese were accompanied by lower carbon contents and were more prone to be populated by oxygen and hydrogen. Manganese is a substitutional element, while oxygen and deuterium are interstitials. Previous reports suggest that the density of segregation sites available for the interstitial segregation is higher than those available for substitutional alloying elements [36]. The oxygen and deuterium segregation widths in Figure 2(c-g) are hence wider than the manganese depletion, which could be ascribed to relatively more segregation sites available for interstitials. This also explains the wider carbon segregation width in Figure 1 compared to manganese. This trend can be further demonstrated schematically by Figure 4c. Consistent changes in the compositional width of GBs along a GB have been reported previously [41]. The inhomogeneous segregation widths evidenced in Figure 2(c-g) can also be ascribed to the faceted structure of GB [24,26,27] and different solute elements may prefer to segregate at different facets or junctions leading to different widths. The degrees of freedom associated with the description of a GB [42,43] are different at different positions of a faceted GB, and all affect the segregation behavior, which can explain the observed variations in the segregation [44]. This also explains the inhomogeneous segregation of carbon at the GB shown in Figure 4 which also appears faceted.

APT revealed hydrogen and oxygen segregation at GBs in a model TWIP steel, which can promote intergranular cracking through HEDE mechanism. GBs depleted in manganese appear more susceptible to oxidation and HE. The current study also evidences that carbon segregation at GBs prevents oxygen and likely hydrogen ingress and protects them from embrittlement. Our study hence motivates further investigations through e.g. atomistic simulations to examine the interdependence of HE and oxidation mechanisms in TWIP steels, while also considering the role of individual alloying elements such as carbon and manganese, to help facilitate the enhanced design and application of these steels.

## Acknowledgements

We thank Uwe Tezins, Christian Bross, and Andreas Sturm for their technical support to the FIB and APT facilities at MPIE. Dr. Leigh T. Stephenson and Dr. Se-Ho Kim are gratefully acknowledged for helping with the cryogenic transfers. H.K. and B.G. acknowledge the financial support from the ERC-CoG-SHINE-771602.

## References

[1] B.C. De Cooman, O. Kwon, K.G. Chin, State-of-the-knowledge on TWIP steel, Mater. Sci. Technol. 28 (2012) 513–527. https://doi.org/10.1179/1743284711Y.0000000095.

[2] M. Koyama, E. Akiyama, Y.K. Lee, D. Raabe, K. Tsuzaki, Overview of hydrogen embrittlement in high-Mn steels, Int. J. Hydrogen Energy. 42 (2017) 12706–12723. https://doi.org/10.1016/j.ijhydene.2017.02.214.

[3] D.M. Bastidas, J. Ress, J. Bosch, U. Martin, Corrosion mechanisms of high-mn twinning-induced plasticity (Twip) steels: A critical review, Metals (Basel). 11 (2021) 1–45. https://doi.org/10.3390/met11020287.

[4] A.E. Salas Reyes, G.A. Guerrero, J.F. Flores álvarez, J.F. Chávez Alcalá, A. Salinas, I.A. Figueroa, G.L. Rodriguez, Influence of the as-cast and cold rolled microstructural conditions over corrosion resistance in an advanced TWIP steel microalloyed with boron, J. Mater. Res. Technol. 9 (2020) 4034–4043. https://doi.org/10.1016/j.jmrt.2020.02.030.

[5] L. Wan, W.T. Geng, A. Ishii, J.P. Du, Q. Mei, N. Ishikawa, H. Kimizuka, S. Ogata, Hydrogen embrittlement controlled by reaction of dislocation with grain boundary in alpha-iron, Int. J. Plast. 112 (2019) 206–219. https://doi.org/10.1016/j.ijplas.2018.08.013.

[6] A. Oudriss, J. Creus, J. Bouhattate, C. Savall, B. Peraudeau, X. Feaugas, The diffusion and trapping of hydrogen along the grain boundaries in polycrystalline nickel, Scr. Mater. 66 (2012) 37–40. https://doi.org/10.1016/j.scriptamat.2011.09.036.

[7] T.M. Harris, M. Latanision, Grain boundary diffusion of hydrogen in nickel, Metall. Trans. A. 22 (1991) 351–355. https://doi.org/10.1007/BF02656803.

[8] A. Saeed-Akbari, J. Imlau, U. Prahl, W. Bleck, Derivation and variation in composition-dependent stacking fault energy maps based on subregular solution model in high-manganese steels, Metall. Mater. Trans. A Phys. Metall. Mater. Sci. 40 (2009) 3076–3090. https://doi.org/10.1007/s11661-009-0050-8.

[9] L. Ismer, T. Hickel, J. Neugebauer, Ab initio study of the solubility and kinetics of hydrogen in austenitic high Mn steels, Phys. Rev. B - Condens. Matter Mater. Phys. 81 (2010) 20–23. https://doi.org/10.1103/PhysRevB.81.094111.

[10] H.K.D.H. Bhadeshia, Prevention of Hydrogen Embrittlement in Steels, ISIJ Int. 56 (2016) 24–36. https://doi.org/10.2355/isijinternational.ISIJINT-2015-430.

[11] Y.J. Kwon, H.J. Seo, J.N. Kim, C.S. Lee, Effect of grain boundary engineering on hydrogen embrittlement in Fe-Mn-C TWIP steel at various strain rates, Corros. Sci. 142 (2018) 213–221. https://doi.org/10.1016/j.corsci.2018.07.028.

[12] H. Fu, W. Wang, X. Chen, G. Pia, J. Li, Grain boundary design based on fractal theory to improve intergranular corrosion resistance of TWIP steels, Mater. Des. 185 (2020) 108253. https://doi.org/10.1016/j.matdes.2019.108253.

[13] Y.J. Kwon, S.P. Jung, B.J. Lee, C.S. Lee, Grain boundary engineering approach to improve hydrogen embrittlement resistance in Fe-Mn-C TWIP steel, Int. J. Hydrogen Energy. 43 (2018) 10129–10140. https://doi.org/10.1016/j.ijhydene.2018.04.048.

[14] H. Khanchandani, R. Rolli, H.C. Schneider, C. Kirchlechner, B. Gault, Hydrogen embrittlement of twinning-induced plasticity steels: Contribution of segregation to twin boundaries, Scr. Mater. 225 (2023) 115187. https://doi.org/10.1016/j.scriptamat.2022.115187.

[15] D.G. Brandon, The Structure of High-Angle Grain Boundaries, Acta Metall. 14 (1966) 1479–1484. https://doi.org/https://doi.org/10.1016/0001-6160(66)90168-4.

[16] A.J. Breen, L.T. Stephenson, B. Sun, Y. Li, O. Kasian, D. Raabe, M. Herbig, B. Gault, Solute hydrogen and deuterium observed at the near atomic scale in high-strength steel, Acta Mater. 188 (2020) 108–120. https://doi.org/10.1016/j.actamat.2020.02.004.

[17] H. Khanchandani, S.-H. Kim, R.S. Varanasi, T. Prithiv, L.T. Stephenson, B. Gault, Hydrogen and deuterium charging of lifted-out specimens for atom probe tomography, Open Res. Eur. 1 (2022) 122. https://doi.org/10.12688/openreseurope.14176.2.

[18] H. Khanchandani, A.A. El-Zoka, S.-H. Kim, U. Tezins, D. Vogel, A. Sturm, D. Raabe, B. Gault, L. Stephenson, Laser-equipped gas reaction chamber for probing environmentally sensitive materials at near atomic scale, PLoS One. (2021) 1–19. https://doi.org/10.1371/journal.pone.0262543.

[19] L.T. Stephenson, A. Szczepaniak, I. Mouton, K.A.K. Rusitzka, A.J. Breen, U. Tezins, A. Sturm, D. Vogel, Y. Chang, P. Kontis, A. Rosenthal, J.D. Shepard, U. Maier, T.F. Kelly, D. Raabe, B. Gault, The Laplace project: An integrated suite for preparing and transferring atom probe samples under cryogenic and UHV conditions, PLoS One. 13 (2018) 1–13. https://doi.org/10.1371/journal.pone.0209211.

[20] S. Zaefferer, G. Habler, Scanning electron microscopy and electron backscatter diffraction, Eur. Mineral. Union Notes Mineral. 16 (2017) 37–95. https://doi.org/10.1180/EMU-notes.16.3.

[21] M.K. Miller, K.F. Russell, K. Thompson, R. Alvis, D.J. Larson, Review of atom probe FIB-based specimen preparation methods, Microsc. Microanal. 13 (2007) 428–436. https://doi.org/10.1017/S1431927607070845.

[22] S.P. Tsai, S.K. Makineni, B. Gault, K. Kawano-Miyata, A. Taniyama, S. Zaefferer, Precipitation formation on ∑5 and ∑7 grain boundaries in 316L stainless steel and their roles on intergranular corrosion, Acta Mater. 210 (2021) 116822. https://doi.org/10.1016/j.actamat.2021.116822.

[23] H. Zhao, L. Huber, W. Lu, N.J. Peter, D. An, F. De Geuser, G. Dehm, D. Ponge, J. Neugebauer, B. Gault, D. Raabe, Interplay of Chemistry and Faceting at Grain Boundaries in a Model Al Alloy, Phys. Rev. Lett. 124 (2020) 1–6. https://doi.org/10.1103/PhysRevLett.124.106102.

[24] X. Zhou, A. Ahmadian, B. Gault, C. Ophus, C.H. Liebscher, G. Dehm, Atomic motifs govern the decoration of grain boundaries by interstitial solutes, (n.d.) 1–39.

[25] A. Ahmadian, D. Scheiber, X. Zhou, B. Gault, R.D. Kamachali, W. Ecker, L. Romaner, G. Dehm, C.H. Liebscher, Interstitial segregation has the potential to mitigate liquid metal embrittlement in iron, (2022). http://arxiv.org/abs/2211.15216.

[26] C.H. Liebscher, A. Stoffers, M. Alam, L. Lymperakis, O. Cojocaru-Mirédin, B. Gault, J.

Neugebauer, G. Dehm, C. Scheu, D. Raabe, Strain-Induced Asymmetric Line Segregation at Faceted Si Grain Boundaries, Phys. Rev. Lett. 121 (2018). https://doi.org/10.1103/PhysRevLett.121.015702.

[27] N.J. Peter, T. Frolov, M.J. Duarte, R. Hadian, C. Ophus, C. Kirchlechner, C.H. Liebscher, G. Dehm, Segregation-Induced Nanofaceting Transition at an Asymmetric Tilt Grain Boundary in Copper, Phys. Rev. Lett. 121 (2018) 255502. https://doi.org/10.1103/PhysRevLett.121.255502.

[28] A. Ahmadian, D. Scheiber, X. Zhou, B. Gault, C.H. Liebscher, L. Romaner, G. Dehm, Aluminum depletion induced by co-segregation of carbon and boron in a bcc-iron grain boundary, Nat. Commun. 12 (2021) 1–11. https://doi.org/10.1038/s41467-021-26197-9.

[29] Z. Yu, P.R. Cantwell, Q. Gao, D. Yin, Y. Zhang, N. Zhou, G.S. Rohrer, M. Widom, J. Luo, M.P. Harmer, Segregation-induced ordered superstructures at general grain boundaries in a nickel-bismuth alloy, Science (80-. ). 358 (2017) 97–101. https://doi.org/10.1126/science.aam8256.

[30] H. Khanchandani, D. Ponge, S. Zaefferer, B. Gault, Materialia Hydrogen-induced hardening of a high-manganese twinning induced plasticity steel, Materialia. 28 (2023) 101776. https://doi.org/10.1016/j.mtla.2023.101776.

[31] X. Zhou, X. Li, K. Lu, Size Dependence of Grain Boundary Migration in Metals under Mechanical Loading, Phys. Rev. Lett. 122 (2019) 1–6. https://doi.org/10.1103/PhysRevLett.122.126101.

[32] O. Barrera, D. Bombac, Y. Chen, T.D. Daff, E. Galindo-Nava, P. Gong, D. Haley, R. Horton, I. Katzarov, J.R. Kermode, C. Liverani, M. Stopher, F. Sweeney, Understanding and mitigating hydrogen embrittlement of steels: a review of experimental, modelling and design progress from atomistic to continuum, J. Mater. Sci. 53 (2018) 6251–6290. https://doi.org/10.1007/s10853-017-1978-5.

[33] S. Ahlawat, K. Srinivasu, A. Biswas, N. Choudhury, First-principle investigation of electronic structures and interactions of foreign interstitial atoms (C, N, B, O) and intrinsic point defects in body- and face-centered cubic iron lattice: A comparative analysis, Comput. Mater. Sci. 170 (2019) 109167. https://doi.org/10.1016/j.commatsci.2019.109167.

[34] T. Hickel, R. Nazarov, E.J. McEniry, G. Leyson, B. Grabowski, J. Neugebauer, Ab initio based understanding of the segregation and diffusion mechanisms of hydrogen in steels, Jom. 66 (2014) 1399–1405. https://doi.org/10.1007/s11837-014-1055-3.

[35] Y.A. Du, L. Ismer, J. Rogal, T. Hickel, J. Neugebauer, R. Drautz, First-principles study on the interaction of H interstitials with grain boundaries in α- and γ-Fe, Phys. Rev. B - Condens. Matter Mater. Phys. 84 (2011). https://doi.org/10.1103/PhysRevB.84.144121.

[36] L. Zhang, B. Radiguet, P. Todeschini, C. Domain, Y. Shen, P. Pareige, Investigation of solute segregation behavior using a correlative EBSD/TKD/APT methodology in a 16MND5 weld, J. Nucl. Mater. 523 (2019) 434–443. https://doi.org/10.1016/j.jnucmat.2019.06.002.

[37] M.P. Seah, INTERFACE, (1979).

[38] Z. Wang, H. Wu, Y. Wu, H. Huang, X. Zhu, Y. Zhang, H. Zhu, X. Yuan, Q. Chen, S. Wang, X. Liu, H. Wang, S. Jiang, M. Kim, Z. Lu, Solving oxygen embrittlement of refractory high-entropy alloy via grain boundary engineering, Mater. Today. 54 (2022) 83–89.

[39] Y. Zhao et al., Chapetr 2-Steel corrosion in concrete, Steel Corros. Concr. Crack. (2016) 19–29.

[40] K.N. Sasidhar, H. Khanchandani, S. Zhang, A. Kwiatkowski, C. Scheu, B. Gault, D. Ponge, D. Raabe, Understanding the protective ability of the native oxide on an Fe-13 at % Cr alloy at

the atomic scale : A combined atom probe and electron microscopy study, 211 (2023).

[41] Z. Peng, T. Meiners, Y. Lu, C.H. Liebscher, A. Kostka, D. Raabe, B. Gault, Quantitative analysis of grain boundary diffusion, segregation and precipitation at a sub-nanometer scale, Acta Mater. 225 (2022) 117522. https://doi.org/10.1016/j.actamat.2021.117522.

[42] A. Stoffers, O. Cojocaru-Mirédin, W. Seifert, S. Zaefferer, S. Riepe, D. Raabe, Grain boundary segregation in multicrystalline silicon: correlative characterization by EBSD, EBIC, and atom probe tomography, Prog. Photovoltaics Res. Appl. 23 (2015) 1742–1753. https://doi.org/10.1002/pip.

[43] D. An, T.A. Griffiths, P. Konijnenberg, S. Mandal, Z. Wang, S. Zaefferer, Correlating the five parameter grain boundary character distribution and the intergranular corrosion behaviour of a stainless steel using 3D orientation microscopy based on mechanical polishing serial sectioning, Acta Mater. 156 (2018) 297–309. https://doi.org/10.1016/j.actamat.2018.06.044.

[44] D. Raabe, M. Herbig, S. Sandlöbes, Y. Li, D. Tytko, M. Kuzmina, D. Ponge, P.P. Choi, Grain boundary segregation engineering in metallic alloys: A pathway to the design of interfaces, Curr. Opin. Solid State Mater. Sci. 18 (2014) 253–261. https://doi.org/10.1016/j.cossms.2014.06.002.

## Supplementary Materials

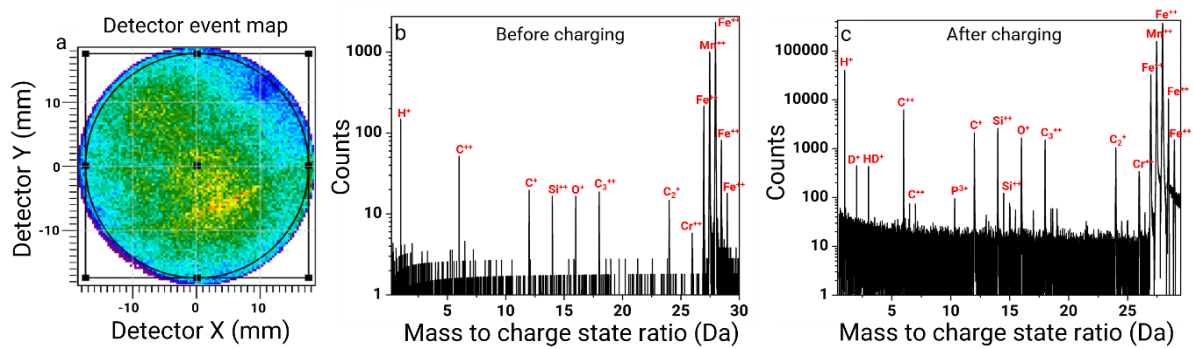

*Supplementary Figure S1. (a) The detector event map corresponding to the pre-charging measurement displaying the first grain boundary (GB) with misorientation angle of 18°. (b) the mass spectrum corresponding to the pre-charging measurement. (c) the mass spectrum corresponding to the analysis following the deuterium charging of the APT specimen reported in Figure 2.*

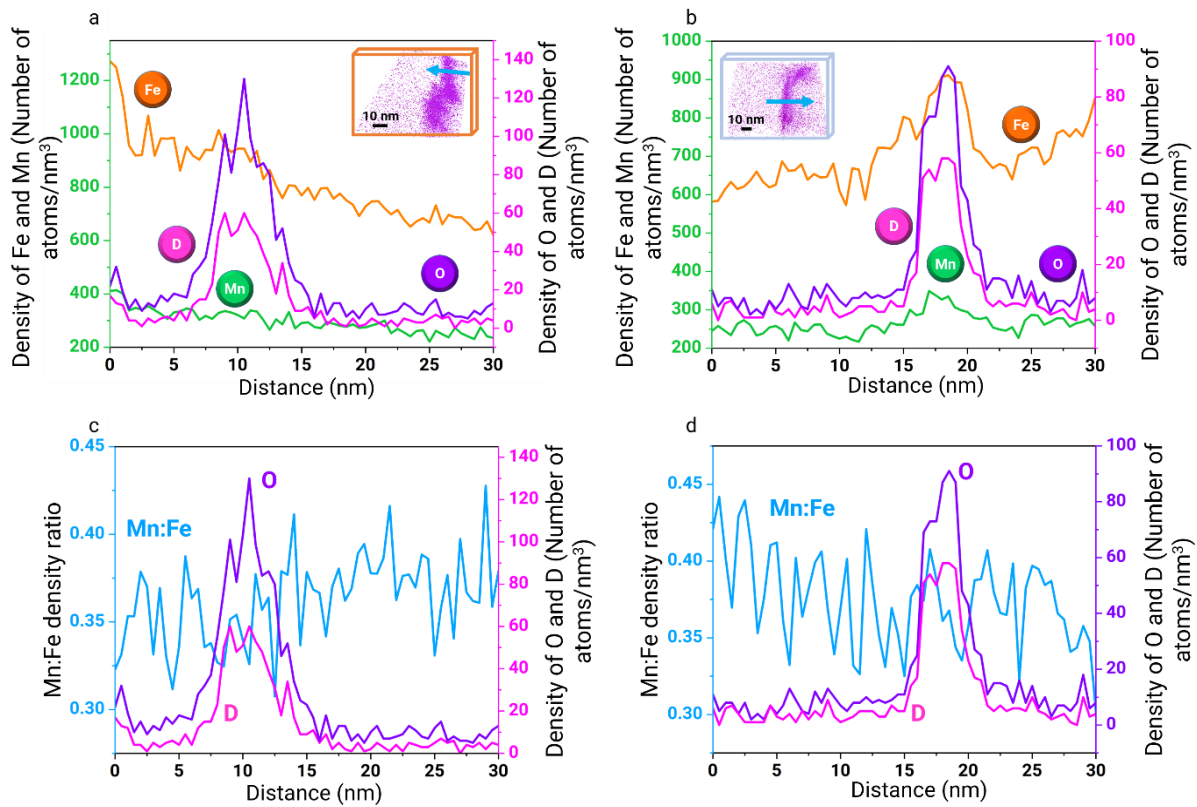

*Supplementary Figure S2. Atomic density profiles for the 1-D composition profiles shown in (a) Figure 2c and (b) Figure 2g; Mn:Fe density ratio plotted with the oxygen and deuterium point densities for (c) the density profile shown in a; and (d) the density profile shown in b.*

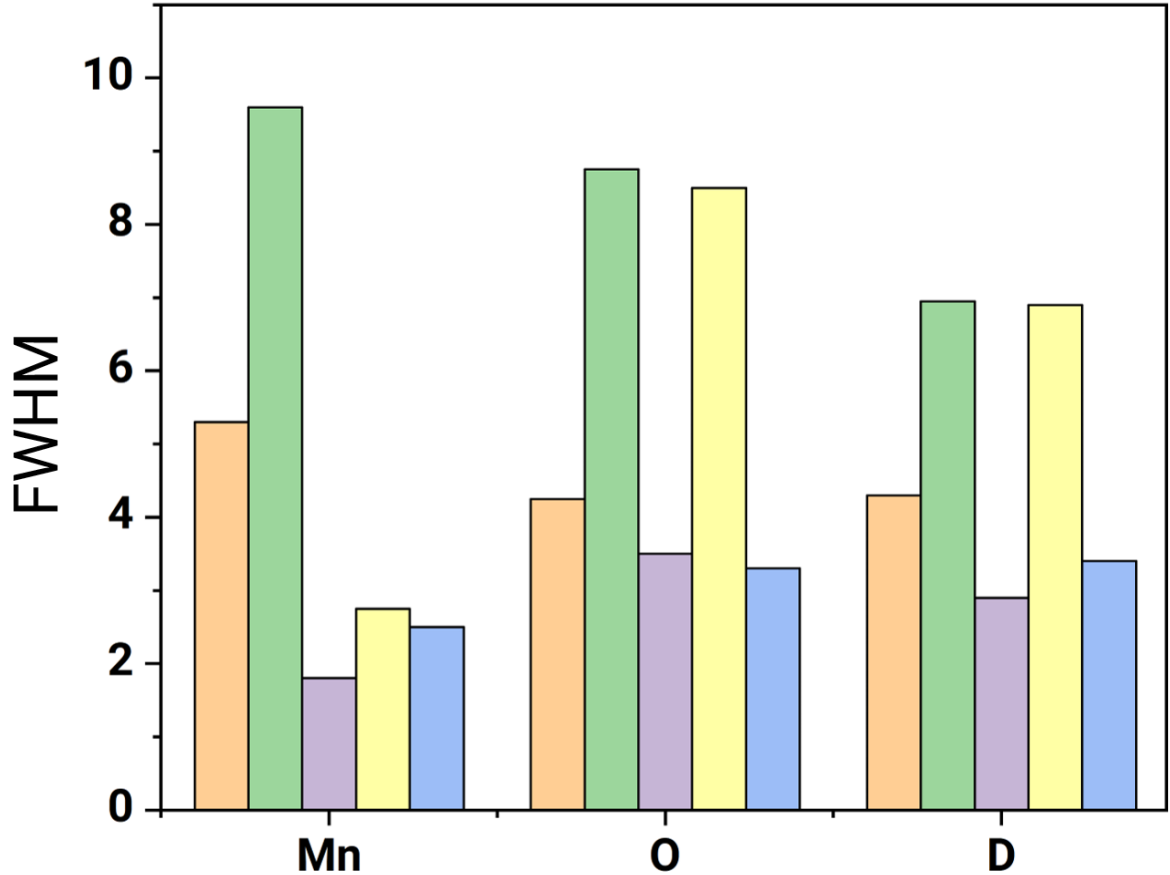

*Supplementary Figure S3. FWHM of manganese, oxygen and deuterium is plotted where each colour corresponds to each profile of Figure 2.*

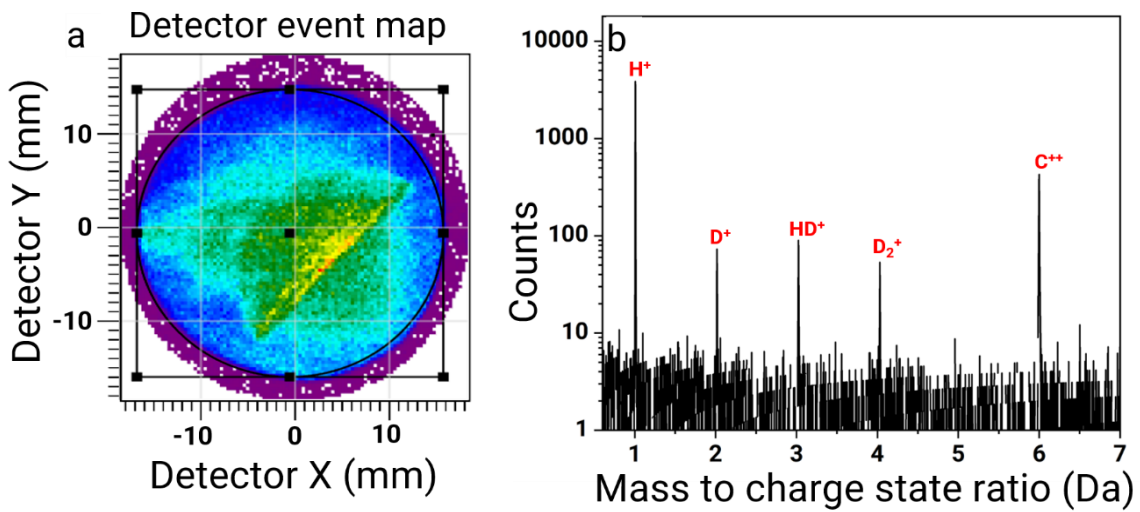

*Supplementary Figure S4. (a) The detector event map confirming the presence of GB reported in Figure 3. (b) The mass spectrum corresponding to the same APT dataset of the deuterium charged GB.*

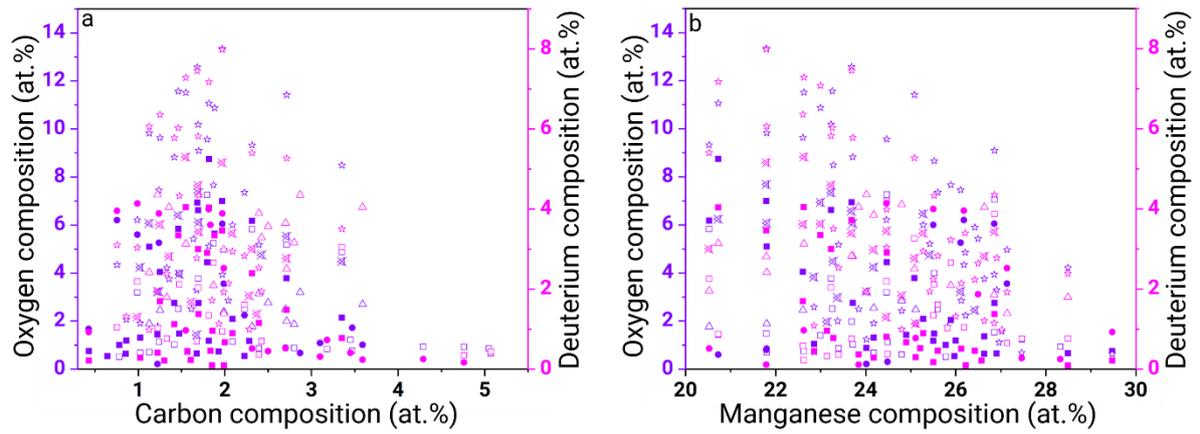

*Supplementary Figure S5. The compositions of oxygen (violet) and deuterium (pink) measured as a function of the composition of (a) carbon and (b) manganese where each symbol represents the data of an individual profile.*

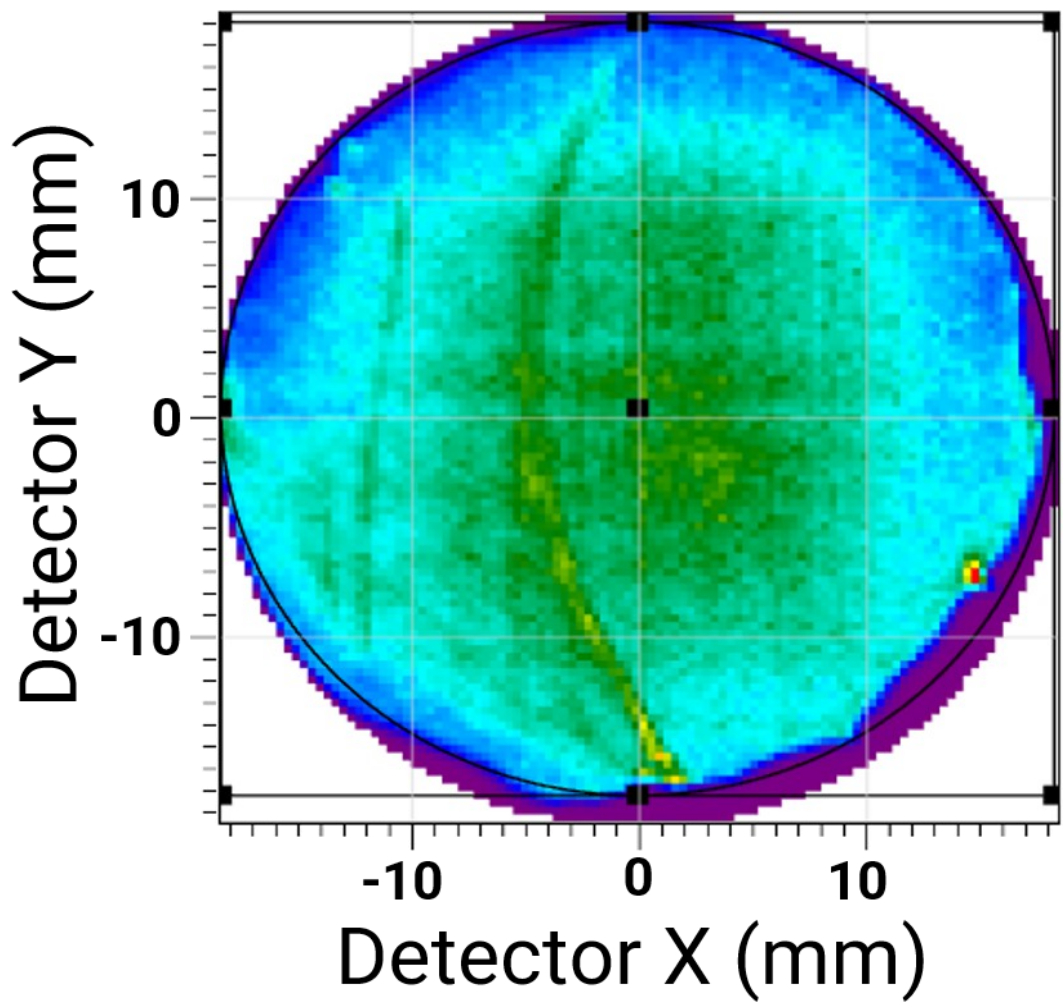

*Supplementary Figure S6. The detector event map confirming the presence of GB reported in Figure 4.*

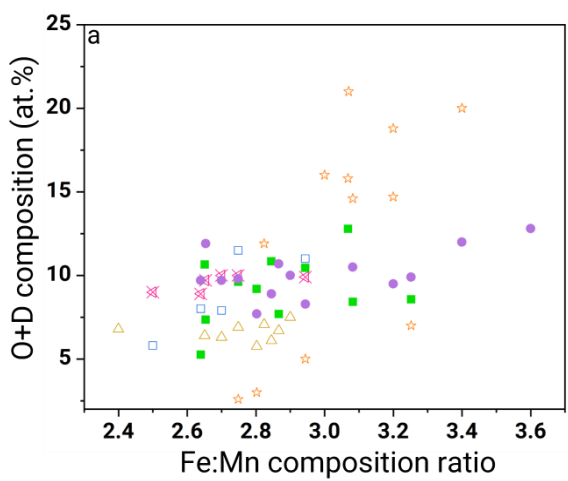
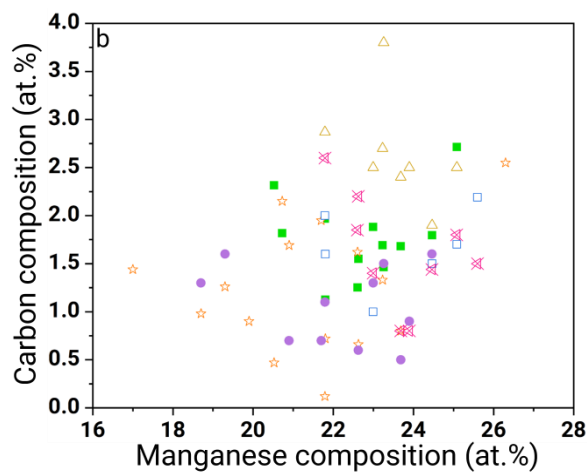

*Supplementary Figure S7. (a) The Fe:Mn composition ratio plotted as a function of oxygen+deuterium compositions. (b) Manganese compositions are plotted vs carbon compositions for each of the analysed GBs. Each symbol represents the data of an individual profile.*